\documentclass[prl,twocolumn,letterpaper, superscriptaddress]{revtex4-1}
\usepackage{graphicx}% Include figure files
\usepackage{dcolumn}% Align table columns on decimal point
\usepackage{bm}% bold math
\usepackage{color}
\usepackage{tabularx}
\usepackage{array}
\usepackage{amsmath}
\usepackage{stmaryrd}  %\llbracket \rrbracket

\begin{document}

\title{Unidirectional Microwave Transduction with Chirality Selected Short-Wavelength Magnon Excitations}

\author{Yi Li}
\email{yili@anl.gov}
\affiliation{Materials Science Division, Argonne National Laboratory, Argonne, IL 60439, USA}

\author{Tzu-Hsiang Lo}
\affiliation{Department of Materials Science and Engineering, UIUC, Urbana, IL 61820, USA}

\author{Jinho Lim}
\affiliation{Department of Materials Science and Engineering, UIUC, Urbana, IL 61820, USA}

\author{John E. Pearson}
\affiliation{Materials Science Division, Argonne National Laboratory, Argonne, IL 60439, USA}

\author{Ralu Divan}
\affiliation{Center for Nanoscale Materials, Argonne National Laboratory, Argonne, IL 60439, USA}

\author{Wei Zhang}
\affiliation{Department of Physics and Astronomy, University of North Carolina, Chapel Hill, NC 27599, USA}

\author{Ulrich Welp}
\affiliation{Materials Science Division, Argonne National Laboratory, Argonne, IL 60439, USA}

\author{Wai-Kwong Kwok}
\affiliation{Materials Science Division, Argonne National Laboratory, Argonne, IL 60439, USA}

\author{Axel Hoffmann}
\email{axelh@illinois.edu}
\affiliation{Department of Materials Science and Engineering, UIUC, Urbana, IL 61820, USA}

\author{Valentine Novosad}
\email{novosad@anl.gov}
\affiliation{Materials Science Division, Argonne National Laboratory, Argonne, IL 60439, USA}

\date{\today}

\begin{abstract}

Nonreciprocal magnon propagation has recently become a highly potential approach of developing chip-embedded microwave isolators for advanced information processing. However, it is challenging to achieve large nonreciprocity in miniaturized magnetic thin-film devices because of the difficulty of distinguishing propagating surface spin waves along the opposite directions when the film thickness is small. In this work, we experimentally realize unidirectional microwave transduction with sub-micron-wavelength propagating magnons in a yttrium iron garnet (YIG) thin film delay line. We achieve a non-decaying isolation of 30 dB with a broad field-tunable band-pass frequency range up to 14 GHz. The large isolation is due to the selection of chiral magnetostatic surface spin waves with the Oersted field generated from the coplanar waveguide antenna. Increasing the geometry ratio between the antenna width and YIG thickness drastically reduces the nonreciprocity and introduces additional magnon transmission bands. Our results pave the way for on-chip microwave isolation and tunable delay line with short-wavelength magnonic excitations.

\end{abstract}

\maketitle

Recent advances in information technologies such as quantum information \cite{BlaisRMP2021}, microelectronics \cite{Razavi2013} and 5G/6G networks \cite{AlFalahyIEEE2017} call for disruptive innovations in microwave signal processing. In particular, circuit-integrated microwave isolators are highly desirable for many applications as they filter unwanted microwave backflow. In quantum information, filtering environmental noise from the output ports is essential for protecting quantum states and entanglement \cite{SuterRMP2016}. Being able to embed isolators on chip will significantly reduce the device volume compared to currently used bulk ferrite-based isolators, and enable non-Hermitian engineering of dynamic systems at a circuit level \cite{GanainyNPhys2017,HurstarXiv2022}.

Magnonics offers opportunities for implementing unidirectional microwave transduction with miniaturized geometry \cite{HarrisJMMM2009,SergaJPD2010,BarmanJPCM2021}. Because of their special dispersion relations \cite{HerringPR1951}, magnons support short-wavelength excitations down to nanometer scale at microwave bandwidth \cite{LiuNComm2018,ShiotaPRL2020,HeinzNanoLett2020,BaumgaertlNanoLett2020,WangAPL2021,CheNComm2022} along with superior frequency tunability with an external magnetic field. In addition, magnons exhibit nonreciprocity based on many unique properties of magnetic excitations, including intrinsic chirality selection in propagating magnetostatic surface spin waves (MSSW) \cite{KhaliliAPL2007,SchneiderPRB2008,SekiguchiAPL2010,JamaliSciRep2013,WongAPL2014,DeoraniCAP2014,NakayamaJJAP2015,ShibataJAP2018}, wavevector-dependent spin wave dispersion shifting \cite{MoonPRB2013,NembachNPhys2015,LanPRX2015,LeeNanoLett2016,BracherPRB2017,GallardoPRApplied2019,VerbaPRApplied2019,ChenPRB2019,WangPRL2020,IshibashiSciAdv2020,HanNanoLett2021}, and non-Hermitian circuit engineering \cite{WangPRL2019,ZhangPRApplied2020}. This allows for compact integration of miniaturized, broad-band and highly tunable isolators in microwave circuits. Furthermore, the realization of miniaturized magnonic isolator is also important for spin wave computing, where magnons are used to carry, transport and process information \cite{ChumakNPhys2015,MahmoudJAP2020}. Unidirectional flow of magnon information enables the isolation of input and output and ensures deterministic logic output \cite{SatoAPExp2013,TalmelliSciAdv2020}.

\begin{figure}[htb]
 \centering
 \includegraphics[width=3.2 in]{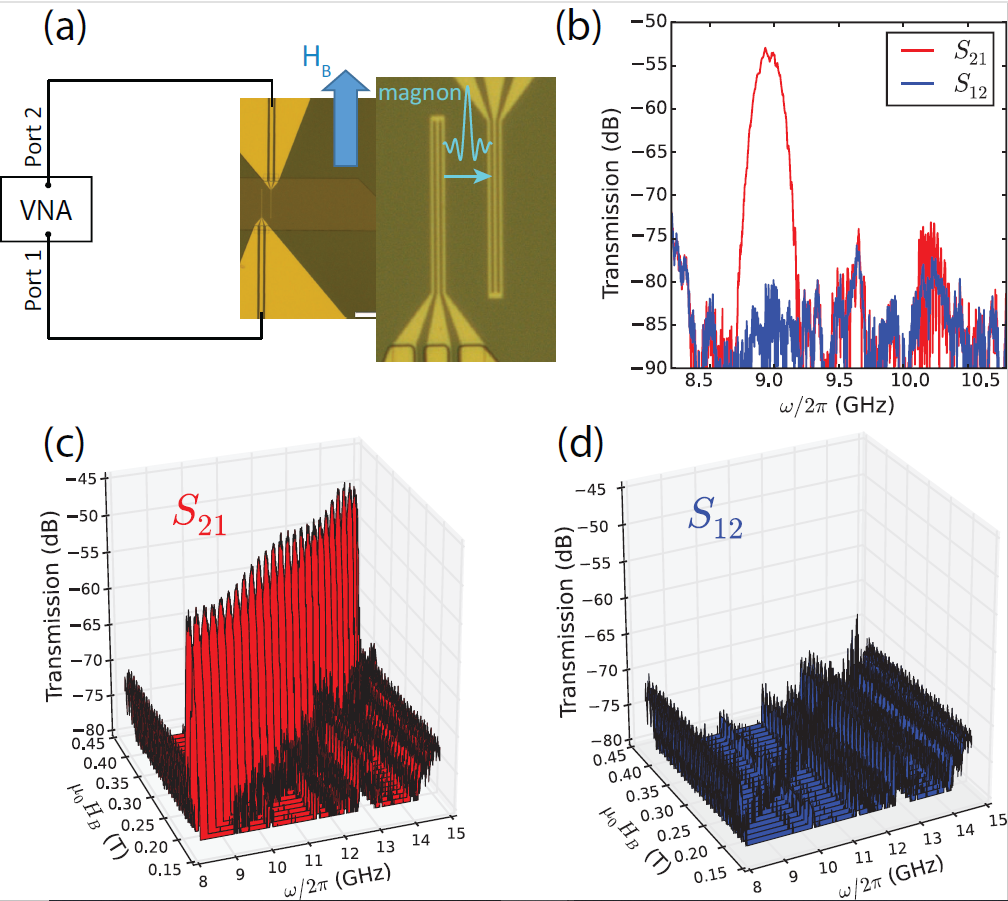}
 \caption{(a) Optical microscope image of the YIG delay line with $t_\text{YIG}=100$ nm, $w=200$ nm and $d=10$ $\mu$m. (b) Nonreciprocal microwave transmission spectra of the device shown in (a) at $\mu_0H_B=0.22$ T, measured by a vector network analyzer. (c-d) Repeated measurements of (b) at different magnetic fields, showing a broad band nonreciprocity with a constant isolation of 30 dB.}
 \label{fig1}
\end{figure}

One major challenge of chip-integrated magnonic microwave isolators is the degradation of nonreciprocity in magnetic thin-film devices. As the film thickness is decreased down to nanometer levels, the $+k$ and $-k$ MSSWs modes, which are localized at the top and bottom surfaces of the film, will permeate to the entire thickness and be both excited by antennas at the top surface, leading to suppressed isolation that are typically less than 10 dB. Other approaches of realizing large magnon nonreciprocity usually suffer from the narrow bandwidth \cite{WangPRL2019,XuSciAdv2020,ShahSciAdv2020}. Alternatively, spin wave nonreciprocity can be achieved by chirality selection from well-designed microwave antennas \cite{DeoraniCAP2014,AuAPL2012,KruglyakAPL2021}, where $+k$ and $-k$ MSSWs exhibit clockwise and counter-clockwise mode profiles depending on the orientation of the magnetization vector. This technique is not restricted by the film thickness and thus can be applied for highly efficient spin wave isolation with proper geometric design.

\begin{figure*}[htb]
 \centering
 \includegraphics[width=6 in]{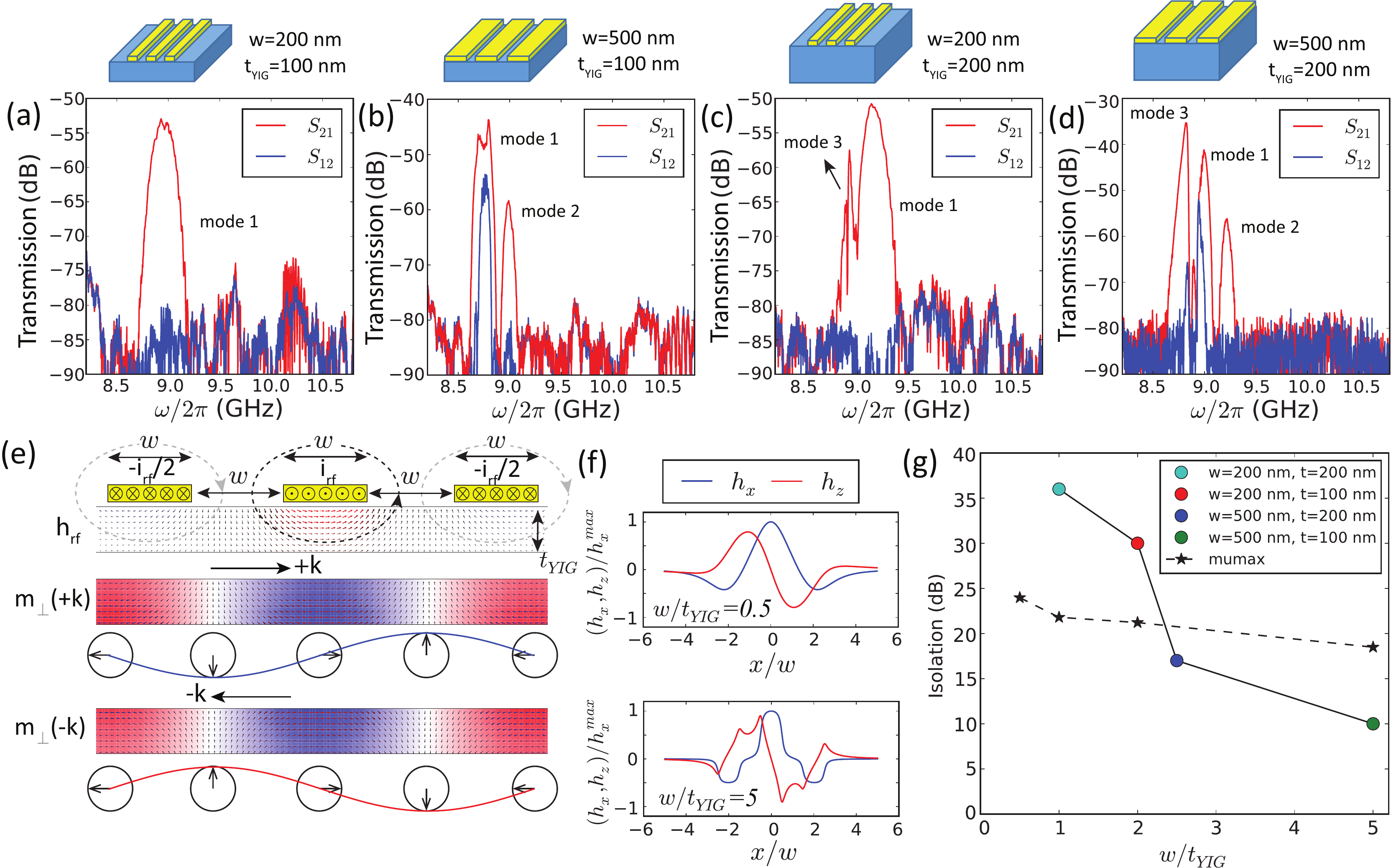}
 \caption{(a-d) Nonreciprocal microwave transmission spectra with different geometries: (a) $w=200$ nm, $t_\text{YIG}=100$ nm; (b) $w=500$ nm, $t_\text{YIG}=100$ nm; (c) $w=200$ nm, $t_\text{YIG}=200$ nm; (d) $w=500$ nm, $t_\text{YIG}=200$ nm. The biasing fields are all $\mu_0H_B=0.22$ T for comparison. (e) OOMMF simulations of the top CPW Oersted field chirality and the $\pm k$ MSSW mode chirality. (f) Oersted field distribution for \textit{top:} $w/t_\text{YIG}=0.5$ and \textit{bottom:} $w/t_\text{YIG}=5$. (g) Summary of experimental magnon isolations as a function of aspect ratio $w/t_\text{YIG}$, along with MuMax simulation results.}
 \label{fig2}
\end{figure*}

In this work, we obtain unidirectional magnon excitations with over 30 dB isolation in yttrium iron garnet (YIG) thin film delay lines with thickness down to 100 nm. We find the leading mechanism of nonreciprocity to be the Oersted field chirality generated from the coplanar waveguide (CPW) antenna, which shows a sinc-function-like spatial profile and selectively couples to the MSSW mode propagating unidirectionally. The isolation quickly decreases from 35 dB to 10 dB as the ratio of the antenna width to the YIG film thickness increases from 1 to 5, which is due to the development of sharp kinks for the Oersted field at the electrode edges. In addition, we also utilize the time-domain functionality of the measurement system to obtain the delay time, group velocity and the nature of the additional microwave transduction bands as the antenna and film geometries deviate from the optimal condition. The demonstration and physical understanding of high-isolation, high-tunability and thin-film-based compact microwave isolator are critical for extending microwave engineering with magnonic devices, and bring new potential for on-chip noise reduction in quantum information applications.

The structure of the delay line is shown in Figure \ref{fig1}(a). A pair of nanofabricated ground-signal-ground (GSG) CPW antennas are patterned on top of a YIG thin-film stripe with a thickness of $t_\text{YIG}=100$ or 200 nm. The three electrodes of the GSG antennas and the gaps between them have an equal width of $w=200$ or $500$ nm and a length of 30 $\mu$m, leading to the excitation of spin waves with a wavelength around $\lambda=4w$ \cite{SushruthPRR2020}. The separation of the two CPWs is $d=5$, 10 or 20 $\mu$m. The external magnetic field is applied parallel to the two CPWs in order to define the MSSW propagation modes between the two antennas.

Figure \ref{fig1}(b) shows the microwave transmission spectra for $t_\text{YIG}=100$ nm, $w=200$ nm and $d=10$ $\mu$m at $\mu_0H_B=0.22$ T, which are measured with a vector network analyzer (VNA). For magnons that propagate from left to right ($S_{21}$) as shown in Fig. \ref{fig1}(a), a single microwave transmission band is observed around 9 GHz and stands well above the microwave background. By reversing the measurement direction from right to left ($S_{12}$), the transmission band disappears. The $S_{21}$ band corresponds to the MSSW modes with the wavevector $k$ orthogonal to the magnetization $M$. By flipping the biasing field, the magnon transmission direction is also reversed \cite{supplement}. With different external biasing fields, the transmission band can be tuned continuously along the frequency axis, as shown in Fig. \ref{fig1}(c). A consistent isolation of 30 dB and a 3-dB bandwidth of 0.15 GHz are measured with negligible change in a broad frequency band from 8 to 15 GHz.

To further explore the role of geometric structure in nonreciprocity, we compare the transmission spectra for a few different antenna geometries, with the results shown in Figs. \ref{fig2}(a-d). Experimentally we see that by increasing $w$ from 200 nm to 500 nm, there is a major increase of the isolated band ($S_{12}$) from below the noise background to being clearly visible, leading to a suppressed nonreciprocity of 15 dB in Fig. \ref{fig2}(b) and 10 dB in Fig. \ref{fig2}(d). In addition, a new side band appears at a higher frequency which is marked as mode 2. This corresponds to the second harmonic of the main MSSW mode.

In order to understand such a large magnon nonreciprocity, we calculate the Oersted field distribution of the GSG antenna as well as the $+k$ and $-k$ MSSW magnon profiles with the experimental geometry. As shown in Fig. \ref{fig2}(e), the Oersted field and the $+k$ mode share the same counter-clockwise spatial chirality, whereas the $-k$ mode exhibit a clockwise chirality \cite{LeeNanoLett2016}. This means the GSG antenna selectively excites the $+k$ mode and is decoupled from the $-k$ mode, leading to chirality induced nonreciprocal magnon excitations. Furthermore, by increasing the aspect ratio between the antenna width and YIG thickness, $w/t_\text{YIG}$, from 0.5 to 5, the Oersted field profile changes from a sinc-like smooth profile to a profile with multiple sharp kinks at the edges of the antenna electrodes [Fig. \ref{fig2}(f)], leading to a finite overlap with the $-k$ mode and a reduction of nonreciprocity. This effect also increases the efficiency of the second harmonic excitation; see the Supplemental Information for more details \cite{supplement}.

We also conduct numerical simulations of the magnon isolations using Mumax in order to compare with experiments. Interestingly, we find a large discrepancy between experiments and simulations, as plotted in Fig. \ref{fig2}(g). The simulations show a slow reduction in isolation from 24 dB to 18 dB as the aspect ratio $w/t_\text{YIG}$ increases from 0.5 to 5. However, the experimental results show a much faster reduction in isolation from 36 dB to 10 dB for $w/t_\text{YIG}$ increasing from 1 to 5. We attribute the disagreement to the different simulation conditions from the experiments as we used a single pixel through the entire thickness to speed up the calculation. This may lead to omission of subtle magnetic profile distributions when the CPW antenna width is comparable to the film thickness. In most previous studies on CPW geometry \cite{DeoraniCAP2014,NakayamaJJAP2015,ShibataJAP2018}, the antenna widths are usually much larger than the film thickness and the chirality selection of the Oersted field has been suppressed. We also note that the skin effect can be ruled out because we do not observe strong frequency dependence of the isolation. The skin depth of gold at 10 GHz is about 785 nm \cite{skindepth}, which is also much larger than the antenna widths of this work so evenly distributed microwave current in the antenna should be a valid assumption.

\begin{figure}[htb]
 \centering
 \includegraphics[width=3.0 in]{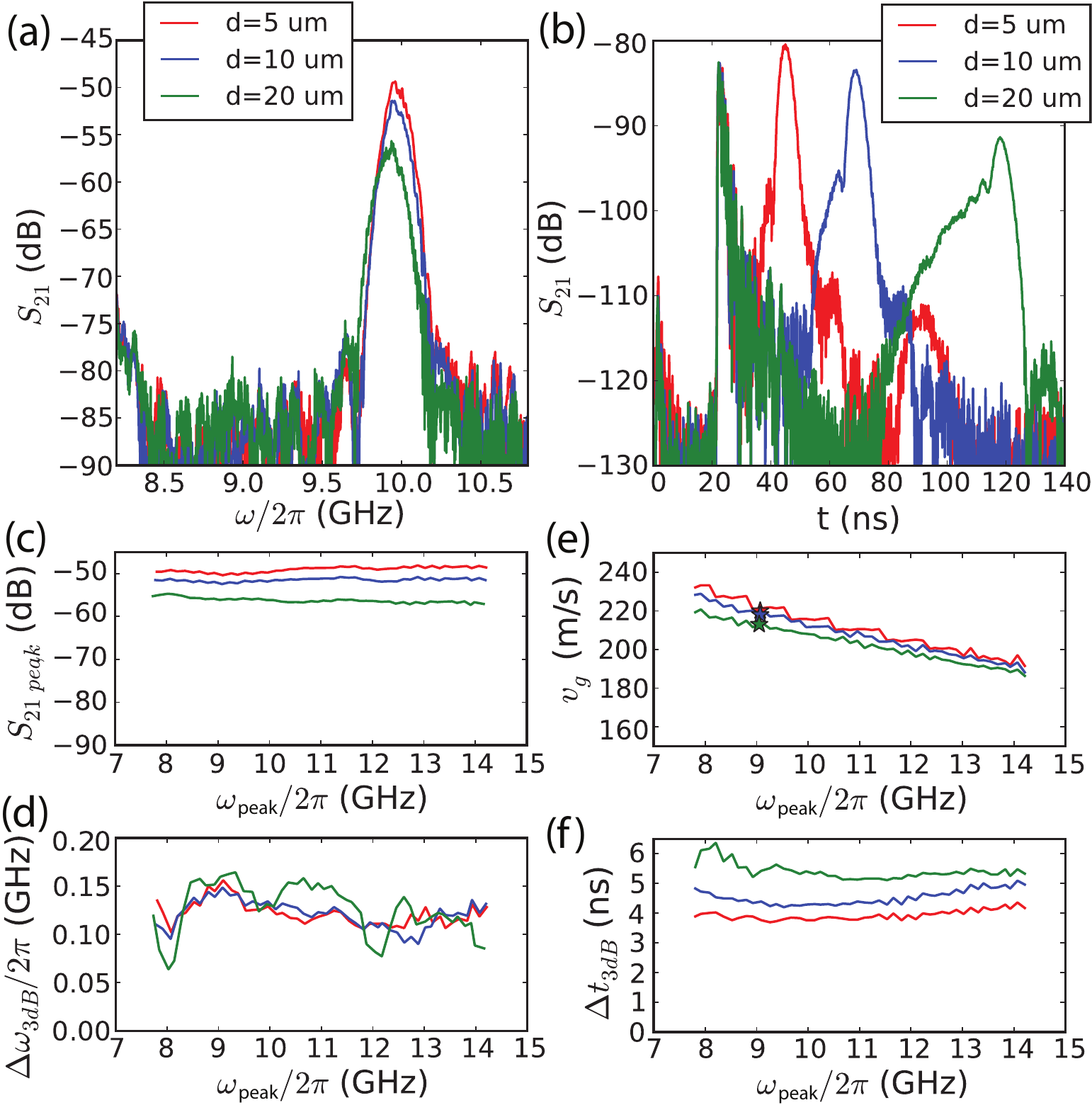}
 \caption{(a) Comparision of microwave transmission spectra with $w=200$ nm, $t_\text{YIG}=100$ nm, for three different antenna separations $d$. (b) IFFT of (a) with a transformation window from 3 to 15 GHz. The biasing fields for (a) and (b) are set as $\mu_0H_B=0.27$ T. (c) Maximal transmission amplitude $S_{21, \text{peak}}$ as a function of the peak position $\omega_\text{peak}$. (d) 3 dB frequency linewidth $\Delta\omega_{3dB}$ as a function of $\omega_\text{peak}$. (c) and (d) are extracted from the measurements of (a) at different fields. (e) Group velocity obtained from the magnon time delay, and plotted as a function of $\omega_\text{peak}$. The stars are the group velocity obtained from the phase changing rate in the frequency spectra measured at $\mu_0H_B=0.22$ T, same as in Fig. \ref{fig1}(b). (f) 3 dB linewidth of the IFFT peak as a function of $\omega_\text{peak}$. (e) and (f) are extracted from the measurements of (b) at different fields. }
 \label{fig3}
\end{figure}

To examine the performance of the nonreciprocal magnonic delay line in the time domain, we also perform time-domain analysis using inverse Fast Fourier Transformation (IFFT) function of the VNA. IFFT simulates the input of a sinc function pulse holding an equal weight of all the frequency components set by the frequency window. As a demonstration, we focus on the delay lines with $w=200$ nm and $t_\text{YIG}=100$ nm. Figs. \ref{fig3}(a) and (b) show the frequency and time domain measurements of the major transmission band ($S_{21}$) at different antenna separations ($d=5$, 10 and 20 $\mu$m). The frequency window for the IFFT analysis of Fig. \ref{fig3}(b) is set to 3 to 15 GHz in order to capture the transmission of the entire frequency band. In the frequency domain, nearly identical transmission bands are measured at different $d$, with only a decreasing transmission amplitude due to the magnon decay with additional propagating distance. In the time domain, the magnon transmission peaks are followed by an initial microwave pulse at 20 ns which is due to the direct microwave radiation between the two antennas serving as the microwave background in \ref{fig3}(a). The time delay of the magnon signal from the microwave radiation pulse scales linearly as a function of $d$, yielding a time delay of 96 ns at $d=20$ $\mu$m or a magnon group velocity of $v_g=208$ m/s at $\omega_\text{peak}/2\pi=10$ GHz. The values of the group velocities are also confirmed from the $S_{21}$ phase changing rate in the frequency domain, as marked by the stars in Fig. \ref{fig3}(e); see the Supplemental Materials for details \cite{supplement}. No magnon signals are measured for the opposite microwave transmission direction ($S_{12}$).

\begin{figure*}[htb]
 \centering
 \includegraphics[width=6.0 in]{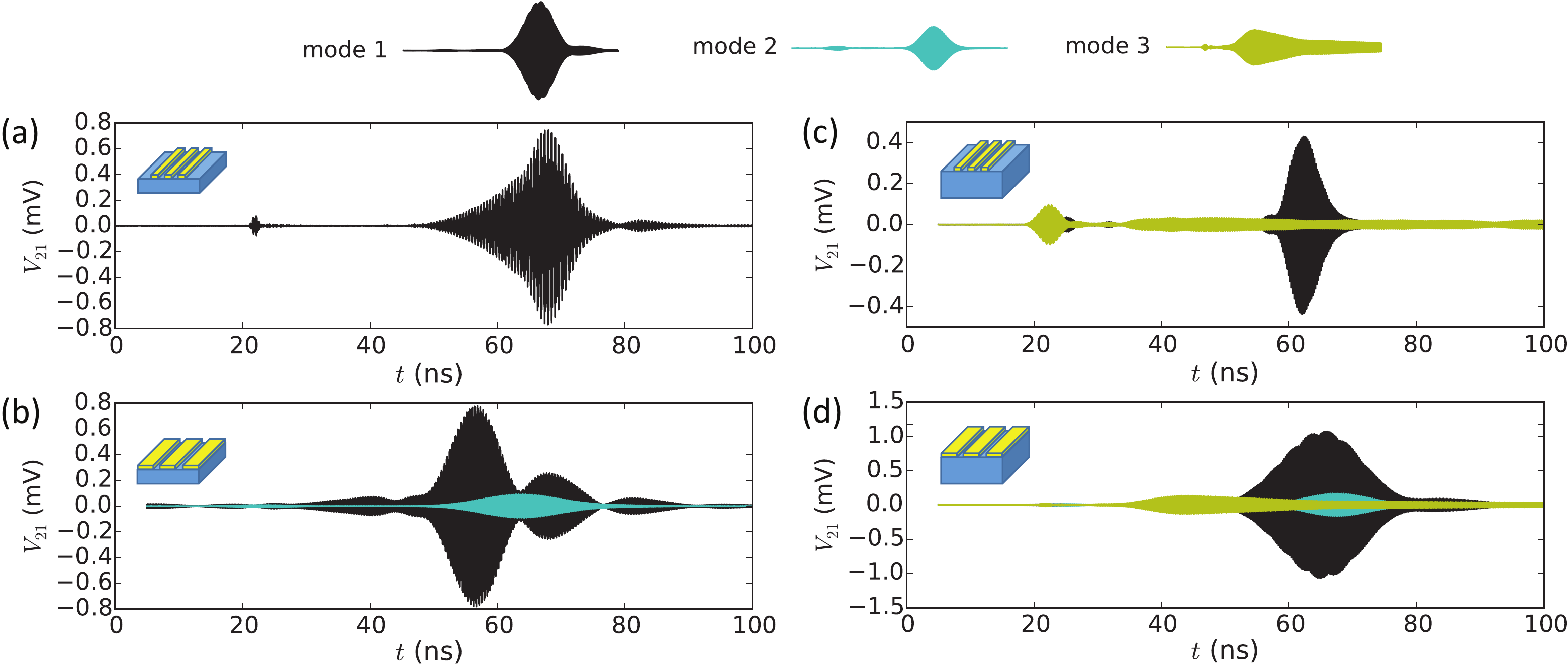}
 \caption{Decomposition of magnonic mode traces with different IFFT frequency windows. (a-d) are converted from IFFT of the frequency domain spectra of Fig. \ref{fig2}(a-d), respectively, with $\mu_0H_B=0.22$ T. (a) Mode 1 is obtained with a frequency window of 8.5-9.5 GHz. (b) Modes 1 and 2 are obtained from frequency ranges as 1: 8.45 to 8.85 GHz, and 2: 8.85 to 9.2 GHz. (c) Modes 1 and 3 are obtained from  frequency ranges as 1: 9.0 to 10 GHz, and 3: 8.5 to 9.0 GHz. (d) Modes 1, 2, 3 are obtained from frequency ranges as 1: 8.88 to 9.1 GHz, 2: 9.1 to 9.6 GHz, and 3: 8.5 to 8.88 GHz. The amplitudes are measured in voltage ($V_{21}$).}
 \label{fig4}
\end{figure*}

The field dependence of the extracted magnon transmission parameters are plotted as a function of peak transmission frequency ($\omega_\text{peak}$) in Figs. \ref{fig3}(c-f). For the frequency domain, all the three devices show nearly frequency-independent peak transmission amplitudes ($S_{21, \text{peak}}$) and linewidth ($\Delta\omega_{3dB}$) from 8 to 14 GHz. The $d$-dependence of $S_{21, \text{peak}}$ yields a magnon decay rate of 0.5 dB/$\mu$m or an exponential decay length of 8.7 $\mu$m. The value of $\Delta\omega_{3dB}/2\pi$ is around 0.12 GHz. Note that it is a function of the GSG antenna geometry and is independent of frequency. For the time domain, the extracted group velocities from time delays are consistent for the three devices and show a slow decrease with frequency, which is the characteristic of MSSW modes. Comparing with analytical calculations, the group velocity values fall between the dipolar and exchange regimes \cite{supplement}, which is in accordance with the range of sub-micron magnon wavelengths.

IFFT analysis also allow us to access the magnon pulse width ($\Delta t_\text{3dB}$), which are in the range of 4-6 ns for the three devices as shown in Fig. \ref{fig3}(f). Because the simulated sinc pulse has a much smaller width ($< 0.1$ ns for a frequency window between 3 and 15 GHz), the measured values of $\Delta t_\text{3dB}$ pose a fundamental limit of the pulse width of the delay line structure. The main source of broadening is the finite spatial width of the antenna for the magnon pulse to pass by. If we take the width of the entire antenna as $4w=800$ nm and the group velocity $v_g$ as 200 m/s, the total traveling time of the magnon pulse across the antenna is $4w/v_g=4$ ns, which is close to the lower bound of $\Delta t_\text{3dB}$ in Fig. \ref{fig3}(f). The increase of $\Delta t_\text{3dB}$ with $d$ is due to the distribution of $v_g$ within the finite bandwidth, $\Delta\omega_{3dB}$, leading to the expansion of the magnon wave package as it propagates. As an estimate, the additional magnon pulse broadening $\Delta t_k$ can be expressed as $\Delta t_k= (\Delta \omega_\text{3dB}/\omega)t$; see the Supplemental Materials for detailed discussion \cite{supplement}. From Fig. \ref{fig3}(d), $\Delta \omega_\text{3dB}/\omega\approx 1$\%. This yields $\Delta t_k\approx 1$ ns by changing $d$ from 5 nm to 20 nm, which agrees with the change of $\Delta t_\text{3dB}$ in Fig. \ref{fig3}(f).

Finally, we use IFFT analysis to identify the nature of additional bands for different antenna geometries in Figs. \ref{fig2}(a-d), which are marked as modes 1, 2 and 3. Fig. \ref{fig4} shows the time-domain profiles of different modes for each geometry. The frequency windows are limited to the edges of each magnon band in order to filter out the contributions from other bands. Mode 1 is known to be the main transmission band. For mode 2 which appears in Figs. \ref{fig4}(b) and (d), the peak position has a longer delay compared with mode 1. This supports that mode 2 is the second harmonic because for MSSW modes the $\omega$-$k$ dispersion curve softens at higher $k$ and the group velocity becomes smaller, resulting in longer travel time of propagating magnons. Mode 3 which appears in Fig. \ref{fig4}(c) and (d) exhibits a long time span of more than 50 ns. From the $S_{21}$ phase changing rate as discussed in the Supplemental Materials \cite{supplement}, we determine mode 3 to be the obliquely launched backward-volume magnetostatic spin waves (BVMSWs) with canted wavevector due to spin wave diffraction \cite{StigloherPRL2016,VlaminckarXiv2022}. From the phase analysis for Fig. \ref{fig2}(d), the high-frequency edge of mode 3 exhibits a near-zero group velocity due to the flat $\omega$-$k$ dispersion at a certain canted angle, causing the magnon pulse to take long time to be transmitted \cite{ChenPRB1994,Demokritov2001,ChumakAPL2009}. We conclude from the time domain analysis that purer magnon modes arise from narrower GSG antenna widths, which is desirable for minimizing loss and decoherence during the pulse microwave processing with the nonreciprocal magnon delay line.

In summary, we present a systematic study of YIG-thin-film magnon delay lines with broad-band isolation above 30 dB. We identify the source of nonreciprocity as the selective coupling of the chiral Oersted field from the GSG antenna to the propagating MSSW modes in only one direction. Using time-domain IFFT analysis, we determine important parameters of the delay line, including time delay, group velocity, bandwidth and time domain broadening. We also identify the nature of the additional magnon transmission bands as the second harmonic and the canted BVMSWs band, which can be suppressed by limiting the aspect ratio of the antenna and the YIG thickness. Our results show a new promise of nonreciprocal magnonic delay lines for processing pulse microwave signals with excellent noise isolation, which may be implemented as chip-embedded microwave isolators for spin wave computing and quantum information processing. We note that our demonstration of nonreciprocal spin wave propagation can be also combined with the recent work of spin-torque spin wave amplification \cite{BreitbacharXiv2022} for implementing unidirectional magnonic microwave nano-amplifiers.

\textbf{Methods.} The devices were fabricated using YIG thin films grown on Gd$_3$Ga$_5$O$_{12}$ (GGG) substrate by liquid phase epitaxy (LPE), which are commercially available \cite{Matesy}. In the first step, chromium hard masks were lithographically patterned on YIG films and used for etching the YIG delay lines. Ar$^+$ ion milling was used for efficiently etching the YIG layer. The Cr masks can be removed by Cr wet etch without damaging the YIG devices. The edges of the two ends are 45 degree away from the long sides in order to minimize magnon reflection. The width of the YIG stripe is 56 $\mu$m and the lengths of the CPW antennas are 30 $\mu$m. In the second step, nano-CPWs were fabricated on top of the YIG devices using e-beam lithography, with Ti(5 nm)/Au(50 nm) electrodes grown by e-beam evaporation to minimize side walls. In the third step, the extended CPWs were fabricated using photolithography. The electrode thicknesses are Ti(5 nm)/Au(150 nm) for $t_\text{YIG}=100$ nm and Ti(5 nm)/Au(250 nm) for $t_\text{YIG}=200$ nm in order to minimize the impedance mismatch across the side walls of the YIG devices.

\textbf{Acknowledgement.} The work was supported by the U.S. DOE, Office of Science, Basic Energy Sciences, Materials Sciences and Engineering Division, with parts of the manuscript preparation, first-principle calculations, device design, and sample fabrication and characterization supported under contract No. DE-SC0022060. U. W, W.-K. K. and V. N. acknowledge support by the U.S. DOE, Office of Science, Basic Energy Sciences, Materials Sciences and Engineering Division. We acknowledge Jiang-Chao Qian, Zhihao Jiang, Andre Schleife, Wolfgang Pfaff and Jian-Min Zuo from UIUC for their helpful discussions.

%\bibliography{ref_propagating_YIG_magnon}


\end{document}